\documentclass[showpacs,twocolumn,pra]{revtex4}
\usepackage{amsfonts}
\usepackage{amsmath}
\usepackage{amssymb}
\usepackage{graphicx}

\begin{document}

\title{ Quantum tunneling time of a Bose-Einstein condensate
traversing through a laser-induced potential barrier}

\author{Zhenglu Duan$^{1}$, Bixuan Fan$^{1}$, Chun-Hua Yuan$^{1}$, Jing Cheng$^{2}$,
 Shiyao Zhu$^{3}$ and Weiping Zhang$^{1}$\footnotemark }

\affiliation{$^{1}$State Key Laboratory of Precision Spectroscopy,
Department of Physics, East China Normal University, Shanghai
200062, China} \affiliation{$^{2}$Department of Physics, South China
University of Technology, Guangzhou 510640, China}

\affiliation{$^{3}$Department of Physics, Hong Kong Baptist
University, Kowloon Tong, Hong Kong, China}

\begin{abstract}
We theoretically study the effect of atomic nonlinearity on the
tunneling time in the case of an atomic Bose-Einstein condensate
(BEC) traversing the laser-induced potential barrier. The atomic
nonlinearity is controlled to appear only in the region of the
barrier by employing the Feshbach resonance technique to tune
interatomic interaction in the tunneling process. Numerical
simulation shows that the atomic nonlinear effect dramatically
changes the tunneling behavior of the BEC matter wave packet, and
results in the violation of Hartman effect and the occurrence of
negative tunneling time.
\end{abstract}

\maketitle

Quantum tunneling of a wave packet through a potential barrier is
one of the fundamental topic in quantum physics
\cite{Hauge0,Landauer,Winful}. The issue of tunneling time has
attracted a lot of attention for decades since it was first put
forward by Condon \cite{Condon}. In the early 1960's, Hartman
predicted that the tunneling time becomes independent of barrier
length for thick enough barriers, ultimately resulting in unbounded
tunneling velocities \cite{Hartman}. Such a phenomenon, termed as
the Hartman effect later on, seems to imply the superluminal
velocities inside the
barriers and leads to a wide interest in many different fields\cite%
{optical,acoustics,Guo,Frank}.

Mathematically, the quantum tunneling is governed by the
Schr\"{o}dinger equation, which is a linear equation describing the
quantum wave nature of a single particle. So far the Hartman effect
or related topics studied in the literature are limited to the
single-particle linear case. After mid-1990's, there are significant
advancements in the realization of atomic Bose-Einstein condensate
(BEC), a macroscopic quantum mechanical wave packet with nonlinear
behavior due to interatomic interactions. Such a macroscopic
coherent matter wave packet of BEC opens a new window to study the
nonlinear quantum dynamics governed by the nonlinear Schr\"{o}dinger
equation or the mean field Gross-Pitaevskii (GP) equation \cite{GP}.
No doubt, tunneling of a BEC matter wave packet would exhibit
different behaviors compared with the single particle tunneling in
the linear quantum mechanics. In fact, there already exist many
studies on BEC tunneling through different kinds of potentials in
the literature \cite{Milburn,potential,potential2}. However, to our
knowledge, the effect of nonlinear interparticle interaction on the
Hartman effect has not been explored.

\begin{figure}[tbp]
\includegraphics[width=2.6in]{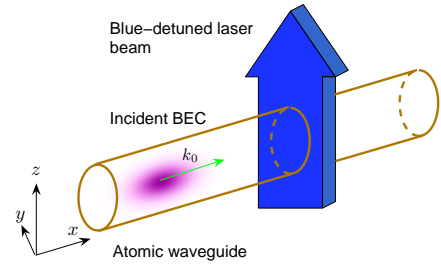}
\caption{{\protect\footnotesize (Color online) Schematic diagram of
a BEC wave packet traversing a blue detuning laser beam in an atomic
waveguide.}} \label{scheme}
\end{figure}

In this brief report, we theoretically investigate how the
interatomic interaction affects the tunneling time in the case of a
coherent BEC wave packet traversing through a potential barrier
created by laser beam. We consider a BEC wave packet confined in a
quasi-one-dimensional atomic waveguide to traverse a potential
barrier created by a far blue-detuned laser beam, which is shown in
Fig. \ref{scheme}. Suppose that the quasi-one-dimensional atomic
waveguide is transversely confined by a harmonic trapping potential,
and the dynamics of atomic BEC can be described by the 3D nonlinear
Schr\"{o}dinger equation
\begin{equation}
i\hbar \frac{\partial }{\partial t}\Psi =\left( -\frac{\hbar ^{2}\nabla ^{2}%
}{2m}+V\left( x\right) +\frac{1}{2}m\omega _{\perp }r_{\perp
}^{2}+gN\left\vert \Psi \right\vert ^{2}\right) \Psi ,
\end{equation}%
where $\Psi $ is the normalized macroscopic wave function of Bose
Einstein condensates, $r_{\perp }=(y,z)$, $m$ the atomic mass, and
$\omega _{\perp }$ the trapping frequency in the radial(transverse)
direction. $g=4\pi \hbar ^{2}a_{s}/m $ describes the interatomic
interaction with $a_{s}$ being the atomic $s$-wave scattering
length. $N$ is the total number of atoms in the condensate, and
$V\left( x\right) $ is the potential barrier created by the
blue-detuned laser beam.

When transverse confinement is very strong, the transverse motion of BEC
atoms may be considered to remain in the ground state. As a result, we can
approximate the total wave function of the BEC as
\begin{equation}
\Psi \left( x,r,t\right) \approx \psi \left( x,t\right) \varphi _{\perp
}\left( r\right) e^{-i\omega _{\perp }t},  \label{total
wavefunction}
\end{equation}%
where $\varphi _{\perp }\left( r\right) $ is the transverse ground state
wave function of the BEC, which can approximately be replaced by a Gaussian
function in the weak nonlinear limit 
\begin{equation}
\varphi _{\perp }\left( r\right) =\frac{1}{\sqrt{\pi }a_{_{\perp }}}\exp
\left( -\frac{r^{2}}{2a_{_{\perp }}^{2}}\right) ,
\label{transverse function}
\end{equation}%
with $a_{_{\perp }}=\sqrt{\hbar /m\omega _{\perp }}$ being the ground state
length of harmonic trapping potential. The longitudinal wave function $\psi
\left( x,t\right) $ is governed by the effective $1$D nonlinear Sch\"{o}%
dinger equation:
\begin{equation}
i\frac{\partial \psi }{\partial t}=-\frac{1}{2}\frac{\partial ^{2}}{\partial
x^{2}}\psi +V\left( x\right) \psi +g\left\vert \psi \right\vert ^{2}\psi .
\label{1 D NLSE}
\end{equation}%
In Eq. (\ref{1 D NLSE}), we express $x$ in units of $a_{_{\perp }}$, $t$ in
units of $\omega _{\perp }^{-1}$, the dimensionless nonlinear interaction $%
g=2a_{s}N/a_{_{\perp }}$, and the dimensionless wave function $\psi =\psi /%
\sqrt{a_{_{\perp }}}$. $V\left( x\right) $ is the potential barrier
experienced by the BEC wave packet and written in the form
\begin{equation}
V\left( x\right) =V_{0}f\left( x\right) ,  \label{V}
\end{equation}%
where $V_{0}$ is the peak value of the potential normalized by $\hbar \omega
_{\perp }$ and $f\left( x\right) $ is the barrier profile, which is
controlled by the laser beam.

Now we have obtained a general dimensionless equation (\ref{1 D
NLSE}) to describe the BEC wave packet transmitting through a
barrier. As pointed out in Ref. \cite{scaling}, Eq. (\ref{1 D NLSE})
is invariant under the scaling transformation $\left\{ x,t,\psi
,V_{0}\right\} \rightarrow \left\{ x\eta ,t\eta ^{2},\psi /\eta
,V_{0}/\eta ^{2}\right\} $ where $\eta $ is a dimensionless
constant. To cover different experimental situations, it is
convenient to introduce such scaling transformation in numerical
simulation. In this paper, we set $\omega _{\perp }\approx 2\pi
\times 100$ Hz and $\eta
=10$, then for $^{87}$Rb atoms, $a_{_{\perp }}\approx \allowbreak 1$ $\mu $%
m. In principle, Eq. (\ref{1 D NLSE}) can describe a BEC wave packet
transmitting through an arbitrary barrier. To simplify the problem,
without loss of physical feature, in this paper we just consider a
rectangular barrier case. Such a rectangular barrier can be
approximately created by a super-Gaussian laser beam with a large
enough order \cite{superG,superG1,HJH}%
. Therefore the barrier profile $f\left( x\right) $ has the form
\begin{equation}
f\left( x\right) =\left\{
\begin{array}{cc}
1, & -\frac{L}{2}<x<\frac{L}{2}, \\
0, & x<-\frac{L}{2}\text{ or }x>\frac{L}{2}.%
\end{array}%
\right.
\end{equation}

In the following, we simulate finite wave packets traversing through
the rectangular barrier via the split operator method \cite{SFT}.
Assume the normalized initial wave packet is Gaussian
\begin{equation}
\psi \left( x,0\right) =\frac{1}{\sqrt{\sqrt{\pi }\Delta x}}\exp \left( -%
\frac{\left( x-x_{0}\right) ^{2}}{2\Delta x^{2}}+ik_{0}\left( x-x_{0}\right)
\right) ,  \label{initial wave packet}
\end{equation}%
where $x_{0}$, $\Delta x$ and $k_{0}$ are the initial center
position, the initial half width and the initial center momentum of
the wave packet, respectively. From Eq. (\ref{1 D NLSE}) we can find
that the interatomic interaction will induce the self-phase
modulation(SPM) \cite{shen,WP Zhanga} of matter wave packet. The SPM
happens in both free region and potential region and causes
confusion of the nonlinear effect on the tunneling process inside
the barrier with the SPM process outside the barrier region. To
avoid this confusion, we shall eliminate the nonlinear interaction
outside of the potential region, i.e., make the $s$-wave scattering
length $a_{s}$ vanishing outside of the potential region. In
principle, this can be realized by employing the Feshbach resonance
technique to tune the atomic scattering length via a spatially
varying magnetic field\cite{WP Zhang}. For spatially varying
magnetic field, the scattering length $a_{s}\left( x\right)
=a_{s0}\left( 1-\Delta B/(B(x)-B_{0})\right) $, where $a_{s0}$ is
the background scattering length,
$\Delta B$ the resonance width, $B_{0}$ the magnetic field of resonance, $%
B(x)$ the magnetic field. By tuning the magnetic field, one can
control the atomic nonlinearity with positive sign (corresponding to
repulsive interatomic interaction), negative one (attractive
interaction) and zero (no interaction). In our work we set the
spatial profile of the scattering length is consistent with that of
the barrier,namely, $g\left( x\right) =g_{0}f(x) $. Considering the
finite width of initial atomic BEC wave packet used in experiment,
we set $\Delta
x=50$ in all the following simulations, corresponding to a half-width $500$ $%
\mu $m for a BEC wave packet composed of $^{87}$Rb atoms.

\begin{figure}[htbp]
\includegraphics[width=2.6in]{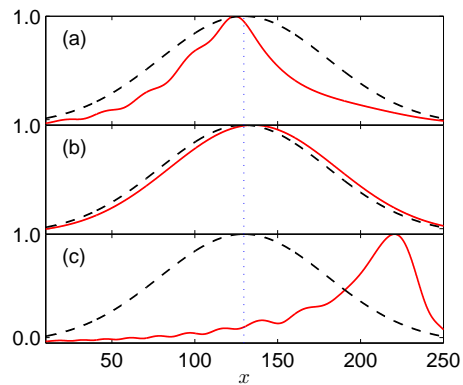}
\caption{{\protect\footnotesize (Color online) The normalized
transmitted wave packets (solid curve) at time $t=440$ with
nonlinear interaction: (a) $g_{0}=-16 $,
(b)$g_{0}=0$ and (c)$g_{0}=200$. Other parameters are $V_{0}=1$, $L=12$ and $%
k_{0}=1.2$. The freely propagating reference wave packet is
represented by the dashed curve. } } \label{evolution}
\end{figure}

With the above assumptions, we now investigate the tunneling of the
BEC wave packet through the barrier with atomic nonlinearity. In
Fig. 2, we present the transmitted wave packets with different
nonlinear interaction at time $t=440$, at which the transmitted wave
packets just emerges from the barrier. Meanwhile we also show the
freely propagating wave packet as reference, for which both the
barrier and nonlinear interaction are set to zero. From the top
panel, one observes that the transmitted wave packet with negative
nonlinearity slightly lags behind the reference one. The middle
panel is the linear case, where the nonlinear interaction is
neglected. In contrast to the negative nonlinear case, the
transmitted wave packet is ahead of the reference one due to the
finite-width effect of incident wave packet, as has been pointed out
by previous works\cite{wavepacket1,wavepacket2,wavepacket3}. In the
bottom panel, we find the transmitted wave packet with positive
nonlinearity is far ahead of the reference one, which is quite
different from other two cases. The results in Fig. 2 show that the
nonlinear interaction indeed, as excepted, affects the tunneling
process of a BEC matter wave packet through the barrier. Below we
explain the physics behind the numerical results in detail.

\begin{figure}[htbp]
\includegraphics[width=2.6in]{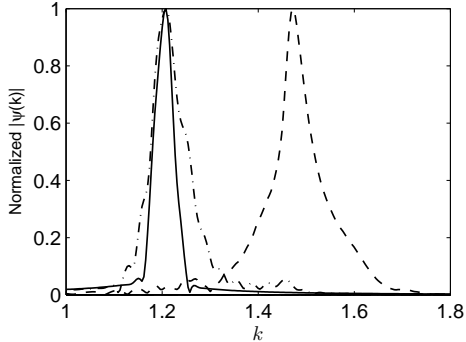}
\caption{{\protect\footnotesize The corresponding momentum spectra
of transmitted wave packets in Fig. \protect\ref{evolution}. Solid
curve stands for the case without nonlinearity, and dashed and
dash-dotted curves for the cases with positive and negative
nonlinearity, respectively.} }
 \label{spectra}
\end{figure}

We consider the transmitted spectra in the momentum space. Due to
the SPM-induced spectral broadening as well as the filtering effect
of the barrier \cite{Winful, Borondo}, the central momenta in the
transmitted spectra exhibit great difference in the linear, positive
and negative nonlinear cases, as shown in Fig. \ref{spectra}. The
filtering of the barrier takes effect on both the linear and
nonlinear cases, while the spectral nonlinear broadening only occurs
in the nonlinear cases. The combination of filtering effect and
nonlinear broadening leads to a large transmitted central momentum
in the positive nonlinear case. Therefore, with the same barrier
parameters and incident wave number, the BEC wave packet with
repulsive interatomic interaction tends to spend less time in the
barrier. However, we note that the transmitted central momenta in
the negative nonlinear and linear cases are less than that in the
positive nonlinear case. This can also be well understood from the
view of point below. In fact, when the interatomic interaction
exists, the atomic nonlinearity modifies the potential barrier and
the atoms "see" an effective potential%
\begin{equation}
V_{eff}=V_{0}f\left( x\right) +g\left( x\right) \left\vert \psi
\left( x,t\right) \right\vert ^{2}.
\end{equation}%
For a repulsive interaction, the nonlinear term has a positive sign.
Consequently, the height of the linear barrier is raised, and vice
versa for an attractive interaction. In terms of the Hartman's
calculation\cite{Hartman}, the wave packet traverses faster through
a higher barrier than a lower one with the same and enough wide
barrier width. Therefore the transmitted wave packet with negative
or positive nonlinearity spend more or less time than the linear one
in the barrier.

Now we turn to the quantitative calculation for the tunneling time
of BEC wave packet through the laser-induced barrier. For a finite
wave packet transmitting through a barrier, we can not use the usual
stationary phase method to find the transmission time of the
transmitted wave packet, because the condition for the stationary
phase method is not satisfied in this situation. However, the widely
used time-of-flight method is applicable to measure the transmission
time of transmitted wave packet \cite{TOA}. The method is described
as follows. Assume that the incident wave packet is placed at the
position $x\left( 0\right)= -x_0(x_0>0)$ at the time $t=0$,
somewhere to the left of the barrier, and let it move to the right
with initial momentum $k_{0}$. After transmitting
through the barrier the position of transmitted wave packet is located in $%
x\left( t_{T}\right) $ at the time $t=t_{T}$. Due to the effect of
barrier and nonlinear interaction, the transmitted wave packet is
usually deformed. In this case, the appropriate way to describe the
center of the transmitted wave packet is to define the expected position as $%
x\left( t_{T}\right) =\int_{x>0}x\left\vert \psi \left( x,t\right)
\right\vert ^{2}dx/\int_{x>0}\left\vert \psi \left( x,t\right) \right\vert
^{2}dx$ \cite{Hauge}. So one gets the tunneling time
\begin{equation}
\Delta t=t_{T}-\frac{x_0 -L/2}{k_{0}}-\frac{x\left( t_{T}\right)
-L/2}{\bar{k}_{0}},  \label{tt}
\end{equation}%
where $\bar{k}_{0}$ is the transmitted central momentum, defined as%
\begin{equation}
\bar{k}_{0}=\int k\left\vert \psi _{T}\left( k\right) \right\vert ^{2}dk,
\end{equation}%
with $\psi _{T}\left( k\right) $ the momentum distribution of
transmitted wave packet.

\begin{figure}[htbp]
\includegraphics[width=2.8in]{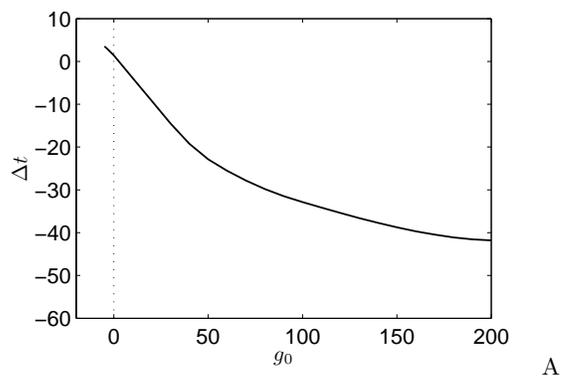}A
\caption{{\protect\footnotesize Tunneling time as a function of the
nonlinear interaction strength. Other parameters are $k_0=0.6$, $V_0=1$ and $L = 6$%
. } } \label{TvsA}
\end{figure}

First, we study the dependence of tunneling time on the nonlinear
interaction. Fig. \ref{TvsA} plots the behavior of tunneling time
via nonlinear interaction strength. Obviously the tunneling time
descends with the nonlinear interaction strength increasing. This
means that the stronger the repulsive interatomic interaction is,
the faster the BEC wave packet traverses through the barrier. It is
interesting to note that the tunneling time is negative for large
positive nonlinearity. The negative tunneling time implies that the
transmitted wave packet exits from the barrier just before the
incident wave packet arrives at the barrier.  Such a phenomenon has
also been found and discussed in the literatures for different
situations\cite{negative time, finite wavepacket1}. Here the
negative tunneling time is due to the repulsive interatomic
interaction, and completely a nonlinear quantum mechanical
phenomenon for an atomic Bose-Einstein condensate.

\begin{figure}[htbp]
\includegraphics[width=2.6in]{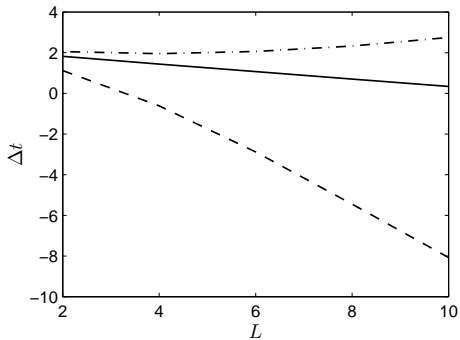}
\caption{{\protect\footnotesize Tunneling time as a function of the
barrier width with nonlinear strength $g_0=0$ (Solid line), $g_0= 5$
(dashed line) and $g_0= -2$ (dash-dotted line). Other parameters are
$k_0 = 0.6$ and $V_0=1$.}} \label{NKL}
\end{figure}

Now, we examine the dependence of tunneling time on the barrier
width as shown in Fig. \ref{NKL}. We find that the tunneling time in
linear and positive nonlinear cases decreases, while increases in
negative nonlinear case, with the barrier width increasing. These
results imply that the Hartman effect is violated in both nonlinear
and linear cases. For the linear case, such a violation is due to
the finite width of the incident wave packet. This has been touched
in many theoretical and experimental works\cite%
{wavepacket1,wavepacket2,wavepacket3}.

In summary, we have numerically studied the effect of nonlinear
interatomic interaction on the quantum tunneling time of a BEC
matter wave packet through a laser-induced barrier. Analysis shows
that both the sign and the strength of nonlinear interaction
significantly affect the tunneling time. As a result, the so-called
Hartman effect in the linear quantum mechanics could be violated in
the nonlinear quantum mechanics with a macroscopic matter wave
packet of Bose-Einstein condensate.

This work is supported by the National Natural Science Foundation of
China under Grant Nos. 10588402, 10774047, and 10874045. the
National Basic Research Program of China (973 Program) under Grant
No. 2006CB921104, the Program of Shanghai Subject Chief Scientist
under Grant No. 08XD14017, Shanghai Leading Academic Discipline
Project under Grant No. B480. We also thank H. Y. Ling and H. Pu for
their helpful discussions.

\end{document}